# Degradation of Co-Evaporated Perovskite Thin Film in Air

Congcong Wang,[1] Youzhen Li,[2] Xuemei Xu,[2] Chenggong Wang,[1] Fangyan Xie,[3] Yongli Gao[1,2*]

[1]Department of Physics and Astronomy, University of Rochester, Rochester, NY, 14627，USA

[2]School of Physics and Electronics, Central South University, Changsha, Hunan, 410083, P. R. China

[3]Instrumental Analysis Center, Sun Yat-Sen University, Guangzhou, 510275, P.R. China

[*]E-mail: ygao@pas.rochester.edu

## Abstract

Methylammonium lead halide perovskites have been developed as highly promising materials to fabricate efficient solar cells in the past few years. The real impact to energy applications relies on the understanding and controlling of the stability of the material. We investigated the degradation of $CH_3NH_3PbI_3$ by air exposure using x-ray diffraction (XRD), x-ray photoelectron spectroscopy (XPS), and atomic force microscopy (AFM). The stoichiometric sample was grown with co-evaporation of $PbI_2$ and $CH_3NH_3I$ on a Au coated Si wafer. It was found that the perovskite thin film gradually turned to $PbI_2$ in air, accompanied with complete removal of N and substantial reduction of I. It was also observed that $PbI_2$ crystallization roughened the film and resulted in a partial exposure of the Au substrate.

Recently, organometal trihalide perovskites have emerged as a new generation of photovoltaic materials with high power conversion efficiency (PCE).[1-6] Among various perovskites, methylammonium lead halide perovskites ($CH_3NH_3PbX_3$, X=Cl, Br, I) have advantages of wide absorption range, high charge-carrier mobility, and low cost.[7-13] The first solid-state perovskite solar cell with a PCE of 9.7% was reported in 2012 by H. S. Kim and co-workers[14]. Ball et al. found that planar heterojunction perovskite solar cell could get around 5% efficiency[15]. This means that perovskite solar cells can be fabricated with lower cost and higher produce efficiency. The PCE was quickly improved to 15.4%[16], and even 19.3% in the following years[17]. As reported, a small area organic-inorganic halide perovskite cell has reached the efficiency of 20.1%.[18,19] Despite the rapid progress in efficiency during past years, perovskite solar cells exhibit significant degradation over a relatively short period of time.[18, 20] Grätzel and co-workers reported that perovskite solar cell could be fabricated under controlled atmospheric conditions with a humidity <1%[21]. Yang and co-workers reported that perovskite solar cell prepared by $PbCl_2$ and $CH_3NH_3I$ in controlled moisture environment could get a good crystal structure and the PCE of 17.1%[22]. The reports all mentioned that the performance of perovskite solar cell was sensitive to moisture. More recently, Niu *et al.* observed the degradation process and proposed that the degradation progressed with $H_2O$ as a catalyst.[23] Kamat *et al.* showed that $H_2O$ is able to react with perovskite in darkness, forming a hydrate product similar to $(CH_3NH_3)_4PbI_6 \cdot 2H_2O$.[24] Kelly *et al.*

suggested the formation of a hydrated intermediate containing isolated $PbI_6^{4-}$ octahedra as the first step of the degradation process.[25] It remains a critical issue to understand this degradation to better control and develop perovskite solar cells that are suitable for renewable energy applications.

In this article, we present our investigations on the degradation of $CH_3NH_3PbI_3$ by air exposure. The evolution of the film was monitored with x-ray diffraction (XRD) at predetermined time intervals while being exposed to air. It was found that the perovskite thin film gradually turned to $PbI_2$, and it took 22 hours to destroy the perovskite completely within the XRD probing depth. X-ray photoelectron spectroscopy (XPS) was used to characterize the composition before and after the air exposure. Atomic force microscopy (AFM) was used to monitor changes of the surface morphology. The XRD, XPS, and AFM provide a comprehensive picture of $CH_3NH_3PbI_3$ degradation, characterized by the complete removal of N, substantial reduction of I, partial oxidation of Pb, and roughening of the film by $PbI_2$ crystallization.

The 60 nm $CH_3NH_3PbI_3$ thin film was grown by co-evaporation of $PbI_2$ and $CH_3NH_3I$ on a Au coated Si wafer. The composition was verified with XPS to be of the atomic ratio (C: N: Pb: I: O= 1.29: 1.07: 1.00: 2.94: 0), very close to the ideal $CH_3NH_3PbI_3$. The film was capped by 2 nm $PbI_2$ to provide some protection against the impact of moisture in air and the XRD scan demonstrated it to be crystalline perovskite $CH_3NH_3PbI_3$. For this study, it is critical to use co-evaporated thin film not only because it provides a superior morphology,[16] but also the surface composition close to the ideal ratio is suitable for surface analysis, a condition hardly reached with the spin cast methods widely used in device fabrications.[8, 9, 12, 13, 26-28]

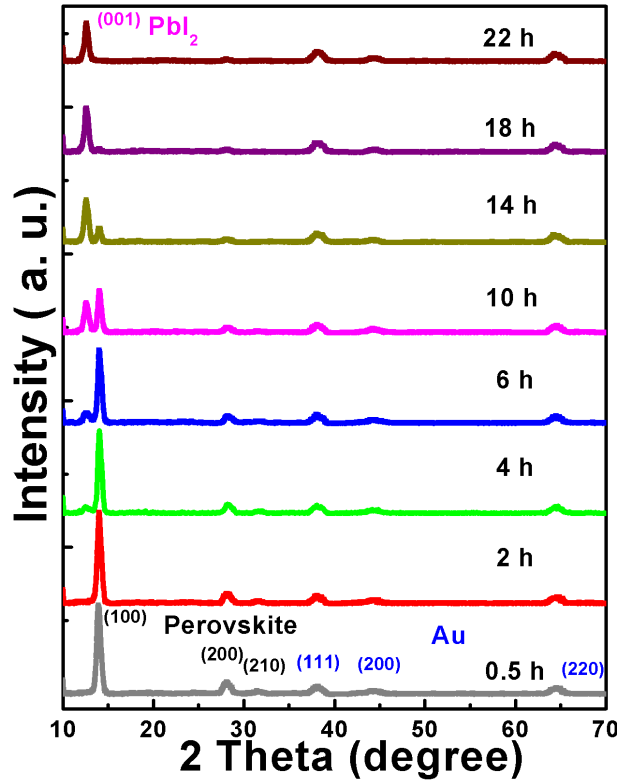

Figure 1. Time evolution of XRD patterns of co-evaporated perovskite film exposed in air. The perovskite features decrease and the PbI$_2$ one increases as the exposure progresses.

Shown in **Figure 1** is the time evolution of XRD patterns of the co-evaporated perovskite film exposed to air. As shown in the bottom pattern, three main perovskite peaks are observed in the pristine sample at 14.02°, 28.20° and 31.52°, assigned to the (100), (200) and (210) lattice planes, respectively. For the region between 14.02° and 28.20°, there are two very small peaks at 21.22° and 24.61° from the (110) and (111) diffraction peaks of CH$_3$NH$_3$PbI$_3$, respectively. These peaks indicate a cubic crystal structure[7]. Other three peaks at 38.19°, 44.38° and 64.58° are the (111), (200) and (220) diffractions of the Au substrate, respectively. It can be seen clearly that the intensities of the three perovskite peaks gradually reduce as the time of air exposure increases, while the peak at 12.68°, corresponding to the (001) diffraction of $PbI_2$, grows during the process. Notably, the film completely converted from $CH_3NH_3PbI_3$ to $PbI_2$ within the XRD probing depth at the angle after exposed to air in about 22 hours.

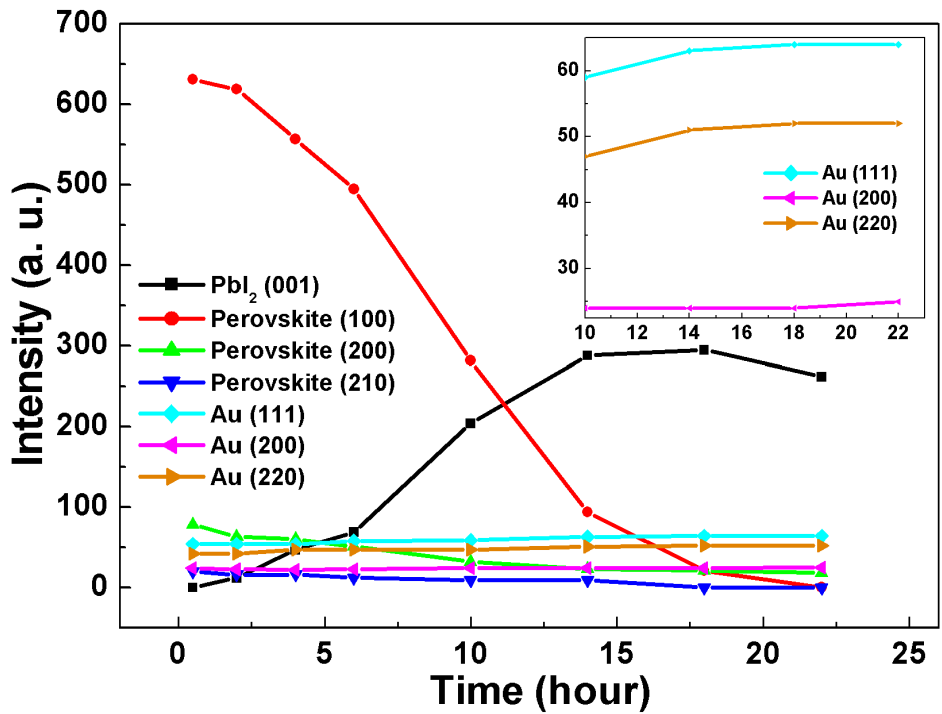

Figure 2. Peak intensity evolution of of $PbI_2$, $CH_3NH_3PbI_3$ and Au. The decrease of the perovskite and the increase of $PbI_2$ are synchronized. The inset shows the small increase of the substrate Au signal at higher exposures, attributed to the roughening of the film by $PbI_2$ crystallization.

The XRD peak heights are plotted in **Figure 2** as a function of air exposure time.

The $CH_3NH_3PbI_3$ (100) peak intensity decreases monotonically to zero. The initial slower rate may be due to the limited protection of the 2 nm $PbI_2$ layer. The $PbI_2$ peak is increasing with exposure time, reciprocal to that of the $CH_3NH_3PbI_3$ one. The rate of the increase drops down after the exposure time $t_{ex}$ of 10 hours. After 18 hours, there is even a decrease of the $PbI_2$ peak intensity. The Au peaks, have a small increase of the intensity, ~18.52% and 23.81%, for the (111) and (200) peak, respectively, after 22 hours of exposure. The inset of the figure is the enlarged evolution of the Au peaks after 10 hours of air exposure. The Au intensities increase following the $PbI_2$ decrease when $t_{ex} > 18$ hours. There may be two possible explanations. First, it may indicate the decrease of the film thickness. Secondly, it may be due to the crystallization of $PbI_2$. As $PbI_2$ crystals aggregate, voids may open in the film and expose the Au substrate.

A more surface sensitive probe like XPS can reveal which of the two possibilities is more likely. **Figure 3** is the XPS full scan of co-evaporated $CH_3NH_3PbI_3$ film before and after air exposure. In Fig. 3(a), there are no obvious other elements except C, N, Pb and I, indicating the uniformity and stoichiometry of the perovskite film. For the air-exposed film, the red marks show the appearance of Au and O peaks, while the blue mark denotes the disappearance of N 1s. This is because of the decomposition of the co-evaporated perovskite into $PbI_2$, as presented in the XRD results, and the absorption of the moisture in the air. The appearance of Au indicates that the film is roughened by air exposure as the perovskite decomposes and $PbI_2$ crystals aggregate. This is because the mean free path of the XPS is only ~2 nm, unable to penetrate the film and see the underlying Au. It can be concluded that it is unlikely the decrease of the film thickness leads to the XRD and the XPS detections of Au in the air-exposed film.

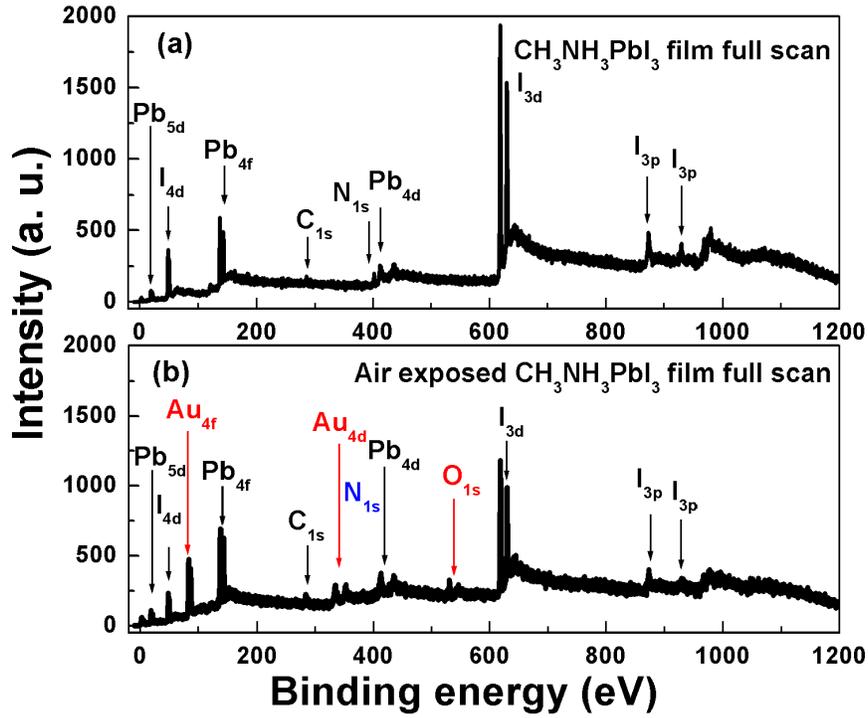

Figure 3. XPS full scan of (a) co-evaporated perovskite film and (b) air-exposed film. The peaks of different elements are all marked. Red marks denote new elements showed in air-exposed film, and the blue mark denotes the element vanished in air-exposed film, compared with the co-evaporated perovskite.

Table 1. Atomic Ratio of co-evaporated film before and after air exposure.

| Element | C | N | Pb | I | O |
|---|---|---|---|---|---|
| Ratio (co-evaporation) | 1.29 | 1.07 | 1.00 | 2.94 | 0 |
| Ratio (after exposure) | 2.22 | 0 | 1.00 | 1.26 | 0.58 |

More critical insight can be obtained from the XPS data on the perovskite decomposition. The atomic ratio comparison from XPS is shown in **Table 1**. All the XPS sensitivity factors and instrumental corrections have been taken into account in obtaining the ratios from the XPS peak areas, and gaussian fitting was used to analyze all peaks. We took the atomic value of Pb as the basis of the atomic ratio as it is the most stable one among the constituencies. For the co-evaporated perovskite, the ratio is C: N: Pb: I: O= 1.29: 1.07: 1.00: 2.94: 0, which is very close to the ideal stoichiometric ratio of $CH_3NH_3PbI_3$ and much superior than those from spin cast ones. [8, 9, 12, 13, 26-28] The C content was a little higher than N and Pb in the pristine co-evaporated film, whose origin may be because of the remaining carbon contamination on the Au substrate. After air exposure, the ratio became C: N: Pb: I: O=2.22: 0:1.00: 1.26: 0.58. Clearly, N was completely disappeared and O increased a lot because of the absorption of moisture in the air. The C increase may be attributed to carbon contamination from the air exposure, as well as the residual amorphous C after the decomposition of the perovskite and the crystallization of $PbI_2$. Iodine decreased during the process as the film transmitted from $CH_3NH_3PbI_3$ to $PbI_2$. However, the ratio 1.26 of I to Pb is less than the stoichiometric value of $PbI_2$. This may be due to the further oxidation of $PbI_2$ during the process and subsequent sublimation of the released iodine. Based on XRD and XPS results, we propose the following process of perovskite degradation in air:

$$CH_3NH_3PbI_3 \xrightarrow{H_2O} C + NH_3(g) + HI(g) + H_2(g) + PbI_2$$

This is consistent to the water catalytic model proposed by Niu et al.,[23] with the notion that the process is irreversible once the gaseous components leave. Furthermore, if oxygen is abundant, the PbI$_2$ may be further oxidized to release iodine, which subsequently sublimes into the ambient and leaves the film iodine deficient and oxygen rich. From the C 1s peak position, the remaining C is amorphous, but it cannot be ruled out that some carbon hydrates were formed during the process, absorbing H$_2$O along the way.

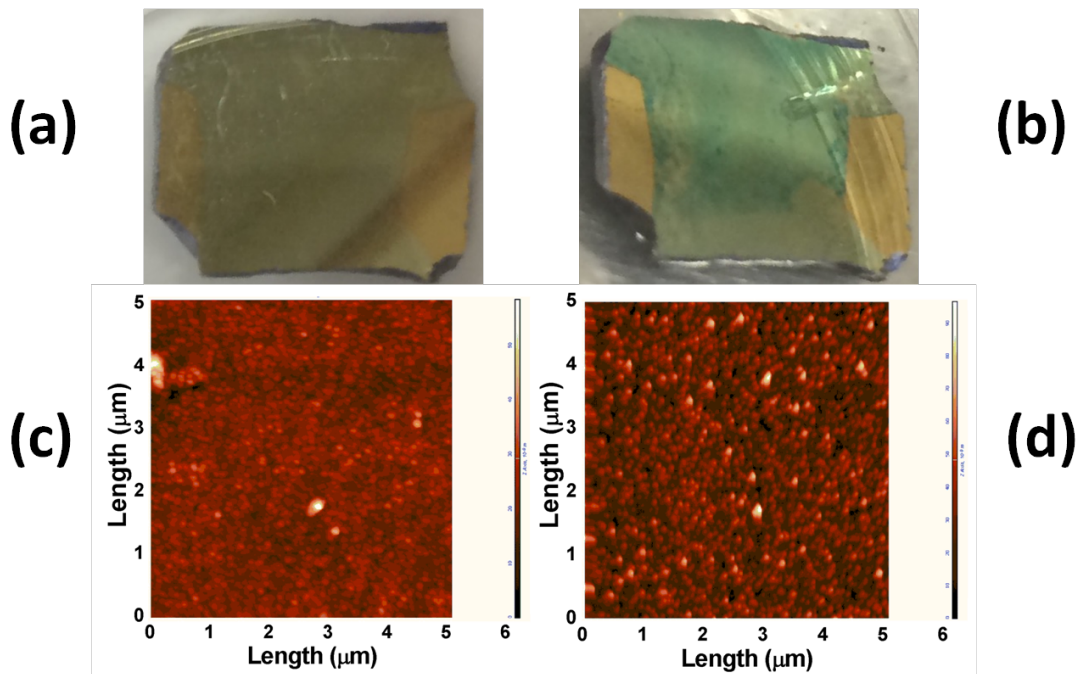

Figure 4. (a) Perovskite film and (b) air exposed film on top of Au/ ITO substrate. (c) AFM image of co-evaporated perovskite. (d) AFM image of air-exposed film.

In **Figure 4**, we compared the morphology of the perovskite film before and after air exposure. When the film was just out of the chamber, it was uniform and the color was light green on top of the Au/ITO substrate as shown in Fig. 4 (a). After exposed in air for

~24h, the film became uneven, and the color of some parts of the film changed to dark green as shown in Fig. 4 (b). According to the XRD results, the light green film is perovskite, whose AFM image is shown in Fig. 4 (c). The uneven dark green film is $PbI_2$, whose AFM image is shown in Fig. 4 (d). Clearly, from the AFM image, the air-exposed film has a lot of voids as we expected from the XRD and XPS results, further confirming the crystallization of $PbI_2$ and the exposure of the Au substrate. The film has a more uniform particle height and smaller particle size before air exposure. It is found that the root mean square (RMS) roughness of the film significantly increases from 3.121 nm to 11.339 nm after the air exposure. The changes of the RMS roughness and particle size confirm our assertion that from $CH_3NH_3PbI_3$ to $PbI_2$ voids are formed with the aggregation of $PbI_2$, causing the partial exposure of the Au substrate.

In conclusion, we investigated the air degradation of stoichiometric perovskite thin film grown by co-evaporation of $PbI_2$ and $CH_3NH_3I$. The XRD measurements revealed that the initial $CH_3NH_3PbI_3$ film transformed to $PbI_2$ after the air exposure. The XPS monitored the composition and electronic structure of the $CH_3NH_3PbI_3$ film before and after the air exposure. The results showed that the moisture in the air catalyzes the decomposition of the perovskite and turns the film into a mixture of amorphous C and crystalline $PbI_2$, releasing gaseous reaction products during the process. The process is therefore irreversible unless made air-tight. The process also results in film roughening and voids creation, demonstrated by the XRD, XPS, and AFM data.

**Experimental section**

*Fabrication conditions:* The co-evaporation process was prepared in a modified Surface Science Laboratories' SSX-100 system. This ultra-high vacuum (UHV) chamber consists of two interconnecting chambers, an evaporation chamber to perform the co-evaporated sample, and an analyzer chamber to do the XPS measurements. The base pressure of the evaporation chamber is typically $1\times10^{-7}$ torr. And it is $1\times10^{-10}$ torr for the analyzer chamber. An X-ray monochromator equipped in the analyzer chamber has a high-throughput bent quartz crystal providing monochromatic Al K$\alpha$ radiation (1486.6 eV). The energy resolution of the XPS is about 0.6 eV. The position of the X-ray spot on the sample can be precisely tuned by the microscope mounted on top of the chamber. And the energy of the photoelectrons was measured by a 40-mm diameter high-resolution detector with parallel detection.

*Materials and Sample Preparation:* $PbI_2$ powder was purchased from Wuhan Jingge Solar Cells limited Company. $CH_3NH_3I$ powder was purchased from Shanghai Zhenpin limited Company (99% purity). The $PbI_2$ and $CH_3NH_3I$ powder were loaded into two boats in the evaporation chamber, and each boat was attached with a thermal couple tightened near the center of the boat to get the evaporation temperature. The substrate is Au coated silicon wafer. It was cleaned with methanol in ultrasound before loaded into the evaporation chamber. After mounting the evaporation sources, the $PbI_2$ and $CH_3NH_3I$ were degassed at the temperature near the evaporation point for about 10 mins. Then, we evaporated $PbI_2$ and $CH_3NH_3I$ independently to get accurate growth rates, respectively.

The film thickness (mass equivalent thickness) was monitored by a quartz crystal microbalance. The $PbI_2$ evaporation rate was kept at ~1.1 Å/min at ~330 ℃, while the $CH_3NH_3I$ evaporation rate was kept at ~1.6 Å/min at ~148 ℃. After the co-evaporation process, the sample was transferred into the analyzer chamber to measure the XPS data without breaking the vacuum. The measurements were all performed at room temperature.

*Characterizations:* The crystalline structure of perovskite film was identified by XRD collected with a Philips APD diffractometer. The XRD diffractometer was equipped with a Cu Kα X-ray tube operated at 40kV and 30 mA using a step size of 0.030 degrees and a time per step of 1.0 s. Our sample was mounted on a low background holder. The measurements started immediately after the film was out of UHV chamber. XRD spectra were measured at 76.4 F, 25% humidity level, and every measurement took about 0.5 h. Experimental fitting of the X-ray data was carried out from 10-70° $2\theta$ at a fixed omega angle of 1 degree. The surface and interface information was measured by XPS with Al Kα radiation. Morphology of co-evaporated and air-exposed film was observed by using the NTMDT AFM Microscope

**Acknowledgments**

The authors thank the National Science Foundation (Grant Nos. CBET-1437656 and DMR-1303742) for financially supporting this research. Y. L and X. X acknowledge supports from National Natrual Science Foundation of China (Grant Nos. 61173047 and 61071025). F. X acknowledges the support from National Natrual Science Foundation of

China (Grant No. 51303217). Technical supports from the XRD Center in Mechanical Department and the Nanocenter in University of Rochester are highly appreciated.